%% file: main.tex
\pdfoutput=1
\documentclass[conference]{IEEEtran}

\usepackage{cite}
\usepackage{amsmath,amssymb,amsfonts}
\usepackage{graphicx}
\usepackage{textcomp}
\usepackage{xcolor}
\usepackage{booktabs}
\usepackage{xspace}
\usepackage{url}
\usepackage[hidelinks]{hyperref}
\usepackage{balance}
\usepackage{tikz}
\usetikzlibrary{arrows.meta,positioning,fit,calc}
\input{macros}

\def\BibTeX{{\rm B\kern-.05em{\sc i\kern-.025em b}\kern-.08em
    T\kern-.1667em\lower.7ex\hbox{E}\kern-.125emX}}

\begin{document}

\title{Observability and Fault Injection for LLM-Based Multi-Agent Systems in Software Engineering}

\author{
\IEEEauthorblockN{
Zahra Seyedghorban\IEEEauthorrefmark{1},
Egor Klimov\IEEEauthorrefmark{2},
Arie van Deursen\IEEEauthorrefmark{1},
Annibale Panichella\IEEEauthorrefmark{1},
Burcu Kulahcioglu Ozkan\IEEEauthorrefmark{1}
}

\IEEEauthorblockA{\IEEEauthorrefmark{1}Delft University of Technology, The Netherlands\\
\{z.seyedghorban, arie.vandeursen, a.panichella, b.ozkan\}@tudelft.nl}

\IEEEauthorblockA{\IEEEauthorrefmark{2}JetBrains Research, The Netherlands\\
egor.klimov@jetbrains.com}
}

\maketitle

\begingroup
\renewcommand{\thefootnote}{}
\footnotetext{\scriptsize
\textcopyright\ 2026 IEEE. Personal use of this material is permitted.
Permission from IEEE must be obtained for all other uses, in any current
or future media, including reprinting/republishing this material for
advertising or promotional purposes, creating new collective works,
for resale or redistribution to servers or lists, or reuse of any
copyrighted component of this work in other works.

\medskip
Published version:
\href{https://doi.org/10.1109/ICST69053.2026.00037}
{https://doi.org/10.1109/ICST69053.2026.00037}
}
\endgroup

\begin{abstract}
Large Language Model-based multi-agent systems are increasingly explored for software engineering tasks, but they remain difficult to inspect, debug, and evaluate under controlled failures. We present \tool, a lightweight and framework-agnostic tool that combines OpenTelemetry-based distributed tracing with fault injection for \llmmas in \se workflows. The tool instruments agent executions with trace-aligned telemetry across workflow phases, agent steps, inter-agent communication, tool calls, and \llm invocations, and supports targeted fault injection at selected interaction points. This makes it possible to compare baseline and faulty executions in a reproducible way and inspect the effects through aligned traces and run artifacts. We describe the motivation, architecture, implementation, current capabilities, and initial validation of the tool on a minimal demo workflow and a real LLM-based multi-agent system for software development.
\end{abstract}


\input{content/01_introduction}
\input{content/02_llmmas_otel}
\input{content/03_validation}
\input{content/04_related_work}
\input{content/05_limitations_future_work}
\input{content/06_tool_availability}
\input{content/07_conclusion}
\input{content/08_acknowledgment}

\enlargethispage{1.5\baselineskip}
\begingroup
\renewcommand{\baselinestretch}{0.9}
\balance
\bibliographystyle{IEEEtran}
\bibliography{bibliography}
\endgroup

\end{document}

%% file: macros.tex
\newcommand{\tool}{\texttt{llmmas-otel}\xspace}
\newcommand{\llm}{LLM\xspace}

\newcommand{\llmmas}{LLM-based multi-agent systems\xspace}

\newcommand{\se}{software engineering\xspace}

\newcommand{\aTwoA}{A2A\xspace}

%% file: content/01_introduction.tex
\section{Introduction}

Large Language Model-based multi-agent systems (LLM-MAS) are increasingly explored for software engineering (SE) tasks such as planning, coding, reviewing, and testing. Instead of relying on a single model call, the work is distributed across multiple specialized agents that communicate, coordinate, and use external tools to complete a task~\cite{qian2024chatdev,phan2024hyperagent}. This makes \llmmas attractive for complex \se workflows, but in practice they still fail frequently and are hard to debug in a principled way \cite{cemri2025mast,deshpande2025trail}. Recent empirical studies report high failure rates across popular MAS frameworks (including SE-oriented ones) \cite{cemri2025mast}.

When these systems fail, the cause is not always obvious. A final incorrect output may come from a much earlier problem, such as a missed constraint, a weak handoff between agents, a tool error, or a model response that quietly pushes the workflow off track. Recent work has shown that failures in multi-agent \llm systems are frequent and varied, including inter-agent misalignment and missing or incomplete verification~\cite{cemri2025mast,ma2025diagnosing}. The difficulty is not only that the workflows are multi-step and distributed across agents, but also that they are stochastic: the same task can unfold differently across runs. As a result, simple logs or final success rates are often not enough to understand what happened or to compare normal and faulty executions in a principled way.

Multi-agent system behavior can be systematically tracked using distributed tracing~\cite{sigelman2010dapper}. A distributed trace captures an entire execution workflow as a single continuous chain, relying on context propagation to link operations together. A trace typically starts from the initial system invocation, such as a user prompt to a planning agent. As the execution context propagates through the system, every distinct operation is recorded as a span. Each span represents a specific part of the execution and contains attributes that provide semantic details about what occurred. OpenTelemetry \cite{opentelemetry} is a vendor-agnostic open standard for observability that defines the conventions for generating and collecting telemetry data, while also providing the SDKs to implement them. To ensure these semantic details are consistent across different instrumentation libraries, OpenTelemetry relies on standard semantic conventions.

Existing work helps us observe that LLM-based multi-agent systems fail, but practitioners still lack a reusable engineering layer for two things: (1) capturing executions in a structured, comparable form across workflow phases, agents, tool calls, and LLM calls, and (2) injecting controlled faults at those same interaction boundaries to study how local perturbations propagate. This paper presents \tool, a lightweight and framework-agnostic tool that addresses this gap through two tightly connected capabilities: observability and fault injection.

A key design choice in \tool is to work \emph{around} an existing LLM-based multi-agent system rather than replacing them. Instead of proposing yet another agent framework, \tool provides instrumentation and injection points that can be integrated into an existing workflow with limited code changes. We report initial validation on both a demo workflow and an existing LLM-MAS. The intended users of \tool are researchers and developers who already have an LLM-based multi-agent workflow and want to inspect its execution or stress-test it under controlled perturbations.
In summary, this paper makes the following contributions:
\begin{itemize}
    \item We present \tool, a reusable tool for observability and fault injection in \llmmas for \se.
    \item We describe a trace model that captures runs across workflow segments, agent steps, communication events, tool invocations, and \llm calls in a consistent and framework-agnostic way.
    \item We provide an initial validation of the tool on a minimal demo workflow and a real LLM-based multi-agent system setting.
\end{itemize}

The rest of the paper is organized as follows. Section~\ref{sec:tool} presents \tool, including its overall design and its observability and fault injection layers. Section~\ref{sec:validation} reports the current validation results. Section~\ref{sec:related} discusses related work and the novelty of the tool. Section~\ref{sec:limitations} outlines current limitations and directions for future work. Finally, Sections~\ref{sec:availability} and~\ref{sec:conclusion} discuss tool availability and conclude the paper.

%% file: content/02_llmmas_otel.tex
\section{Framework}
\label{sec:tool}

\subsection{Overview}

\tool is a lightweight and framework-agnostic tool for observing and perturbing executions of \llmmas in software engineering workflows. Rather than proposing yet another agent framework, it wraps an existing workflow at a small number of meaningful boundaries and produces aligned execution artifacts. At a high level, the tool takes as input an existing LLM-based multi-agent system, a task to execute, and optionally a fault specification. It then produces structured traces together with optional offline message records and trace annotations that make both baseline and faulty runs inspectable in a consistent way.

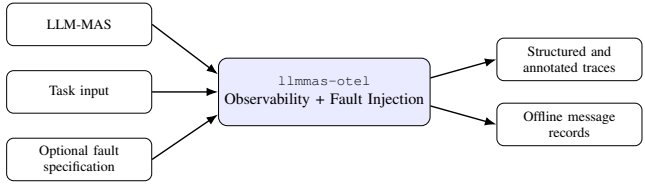
\begin{figure}[t]
\centering
\resizebox{0.95\columnwidth}{!}{%
\begin{tikzpicture}[
    font=\footnotesize,
    >=Latex,
    box/.style={
        draw,
        rounded corners=4pt,
        align=center,
        minimum width=2.9cm,
        minimum height=0.85cm
    },
    core/.style={
        draw,
        rounded corners=5pt,
        fill=blue!8,
        align=center,
        minimum width=3.8cm,
        minimum height=1.35cm
    },
    flow/.style={->, thick}
]

\node[box] (mas)    at (0, 1.35) {LLM-MAS};
\node[box] (task)   at (0, 0)    {Task input};
\node[box] (faults) at (0,-1.35) {Optional fault\\specification};

\node[core] (tool) at (4.9,0) {\tool\\[0.6mm]\small Observability + Fault Injection};

\node[box] (traces)  at (9.8, 0.65) {Structured and\\annotated traces};
\node[box] (records) at (9.8,-0.65) {Offline message\\records};

\coordinate (toolin1)  at ([yshift= 0.35cm]tool.west);
\coordinate (toolin2)  at (tool.west);
\coordinate (toolin3)  at ([yshift=-0.45cm]tool.west);

\coordinate (toolout1) at ([yshift= 0.28cm]tool.east);
\coordinate (toolout2) at ([yshift=-0.28cm]tool.east);

\draw[flow] (mas.east) -- (toolin1);
\draw[flow] (task.east) -- (toolin2);
\draw[flow] (faults.east) -- (toolin3);

\draw[flow] (toolout1) -- (traces.west);
\draw[flow] (toolout2) -- (records.west);

\end{tikzpicture}%
}
\caption{High-level input/output view of \tool. The framework wraps an existing LLM-based multi-agent system, optionally applies configured faults, and produces aligned artifacts for analysis.}
\label{fig:tool_io}
\end{figure}

Figure~\ref{fig:tool_io} shows this high-level view. The underlying LLM-based multi-agent system remains responsible for performing the task itself. \tool is responsible for instrumentation, optional perturbation, and export of aligned artifacts for later analysis.

From a user perspective, adopting \tool consists of three practical steps. First, the user marks a small number of existing workflow boundaries, such as task/session start, workflow phases, agent steps, and selected A2A, tool, or LLM operations. Second, the user runs the target workflow once without perturbation to obtain a baseline trace and optional offline message records. Third, the user optionally enables one or more configured fault rules and re-executes the same task to obtain a structurally aligned faulty run.

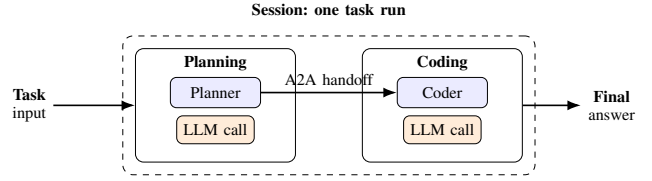
\begin{figure}[t]
\centering
\resizebox{0.95\columnwidth}{!}{%
\begin{tikzpicture}[
    font=\footnotesize,
    >=Latex,
    segment/.style={draw, rounded corners=4pt, minimum width=2.75cm, minimum height=1.95cm},
    agentbox/.style={draw, rounded corners=3pt, fill=blue!8, minimum width=1.55cm, minimum height=0.52cm, align=center},
    op/.style={draw, rounded corners=3pt, fill=orange!15, minimum width=1.25cm, minimum height=0.48cm, align=center}
]

\node[align=center] (input) at (0,0) {\textbf{Task}\\input};

\node[segment] (planseg) at (3.2,0) {};
\node at ($(planseg.north)+(0,-0.25)$) {\textbf{Planning}};
\node[agentbox] (planner) at ($(planseg.center)+(0,0.22)$) {Planner};
\node[op] (planllm) at ($(planseg.center)+(0,-0.42)$) {\llm call};

\node[segment] (codeseg) at (7.1,0) {};
\node at ($(codeseg.north)+(0,-0.25)$) {\textbf{Coding}};
\node[agentbox] (coder) at ($(codeseg.center)+(0,0.22)$) {Coder};
\node[op] (codellm) at ($(codeseg.center)+(0,-0.42)$) {\llm call};

\draw[->, thick] (input.east) -- (planseg.west);
\draw[->, thick] (planner.east) -- node[above, align=center] {\footnotesize A2A handoff} (coder.west);
\draw[->, thick] (codeseg.east) -- ++(1.0,0) node[right, align=center] {\textbf{Final}\\answer};

\node[draw, dashed, rounded corners=5pt, fit=(planseg)(codeseg), inner sep=6pt] (sessionbox) {};
\node at ($(sessionbox.north)+(0,0.45)$) {\textbf{Session: one task run}};

\end{tikzpicture}%
}
\caption{Running example used throughout the paper: a minimal Planner$\rightarrow$Coder LLM-based multi-agent system workflow for a software engineering task.}
\label{fig:demo_workflow}
\end{figure}

Throughout the rest of this section, we use the minimal demo workflow in Figure~\ref{fig:demo_workflow} as a running example. The demo contains the core ingredients of an LLM-based multi-agent system execution. A task is first handled by a \emph{Planner} agent in a \emph{planning} phase. The Planner performs one \llm call to generate a short implementation plan and then sends that plan as an agent-to-agent message to a \emph{Coder}. The workflow then enters a \emph{coding} phase, where the Coder receives the message and performs a second \llm call to produce the final answer. This gives us a compact pipeline with two phases, two agent steps, one inter-agent handoff, and two \llm calls.

\subsection{Observability Layer}

The usefulness of this trace model comes not only from its hierarchy, but also from the semantic information associated with each span. This information is captured through a consistent vocabulary of span attributes, where each attribute has a stable name and meaning across executions. Table~\ref{tab:obs_attributes} summarizes the vocabulary used by the current implementation. This consistency allows traces to be compared and queried systematically across different runs and workflows.

The running example in Figure~\ref{fig:demo_workflow} makes these attributes concrete. At the highest level, the full execution is represented by a \emph{session span}. Inside it, the planning phase becomes a \emph{segment span} with its own name and ordering information, and the Planner's work becomes an \emph{agent\_step span} that records the acting agent and step index. The Planner's model invocation is then captured as an \texttt{llm\_call} span, while the handoff from Planner to Coder is captured through \texttt{a2a\_send} and \texttt{a2a\_receive} spans carrying the communication metadata listed in Table~\ref{tab:obs_attributes}. This means that the trace not only shows that an LLM call occurred or that a message was sent, but also which phase it occurred in, which agent was responsible, and how later steps relate to it.

\begin{table}[t]
\centering
\caption{Main span attributes used by the observability layer.}
\label{tab:obs_attributes}
\setlength{\tabcolsep}{4pt}
\renewcommand{\arraystretch}{1.08}
\begin{tabular}{p{1.15cm} p{3.5cm} p{2.9cm}}
\toprule
\textbf{Span} & \textbf{Attributes} & \textbf{Meaning} \\
\midrule

Session &
\begin{tabular}[t]{@{}l@{}}
{\scriptsize\texttt{session.id}}
\end{tabular}
&
{\scriptsize
Identifies one full task run and groups all child spans under the same execution.
}
\\
\midrule

Segment &
\begin{tabular}[t]{@{}l@{}}
{\scriptsize\texttt{segment.name}}\\
{\scriptsize\texttt{segment.order}}\\
{\scriptsize\texttt{segment.origin (opt.)}}
\end{tabular}
&
{\scriptsize
Records the workflow phase, its position in the run, and optionally where the segment originated from.
}
\\
\midrule

Agent-step &
\begin{tabular}[t]{@{}l@{}}
{\scriptsize\texttt{agent.id}}\\
{\scriptsize\texttt{step.index}}
\end{tabular}
&
{\scriptsize
Identifies which agent performed the step and the turn index of that agent within the run.
}
\\
\midrule

A2A send / receive &
\begin{tabular}[t]{@{}l@{}}
{\scriptsize\texttt{source\_agent.id}}\\
{\scriptsize\texttt{target\_agent.id}}\\
{\scriptsize\texttt{edge.id}}\\
{\scriptsize\texttt{message.id}}\\
{\scriptsize\texttt{channel}}\\
{\scriptsize\texttt{message.preview}}\\
{\scriptsize\texttt{message.sha256}}
\end{tabular}
&
{\scriptsize
Describes the communication edge, the message identity, the channel, and lightweight message-content hints for inspection or matching.
}
\\
\midrule

Tool call &
\begin{tabular}[t]{@{}l@{}}
{\scriptsize\texttt{operation.name}}\\
{\scriptsize\texttt{tool.name}}\\
{\scriptsize\texttt{tool.type (opt.)}}\\
{\scriptsize\texttt{tool.call.id}}\\
{\scriptsize\texttt{tool.args.preview (opt.)}}\\
{\scriptsize\texttt{tool.args.sha256 (opt.)}}
\end{tabular}
&
{\scriptsize
Records which tool was invoked, the call identity, and optional lightweight summaries of tool arguments.
}
\\
\midrule

LLM call &
\begin{tabular}[t]{@{}l@{}}
{\scriptsize\texttt{operation.name}}\\
{\scriptsize\texttt{provider.name}}\\
{\scriptsize\texttt{request.model}}\\
{\scriptsize\texttt{request.id}}\\
{\scriptsize\texttt{llm.input.preview (opt.)}}\\
{\scriptsize\texttt{llm.input.sha256 (opt.)}}
\end{tabular}
&
{\scriptsize
Records the provider, model name, and request identity, plus optional lightweight summaries of prompt input.
}
\\
\midrule

Injected span &
\begin{tabular}[t]{@{}l@{}}
{\scriptsize\texttt{fault.injected}}\\
{\scriptsize\texttt{fault.type}}\\
{\scriptsize\texttt{fault.spec\_id}}\\
{\scriptsize\texttt{fault.decision}}
\end{tabular}
&
{\scriptsize
Marks that a configured fault was applied on an A2A, tool, or LLM boundary and records the decision taken.
}
\\

\bottomrule
\end{tabular}
\end{table}

From the user side, the instrumentation is small and lightweight in the sense that users annotate only a small number of existing workflow boundaries and reuse the host MAS implementation. Users mark a few existing workflow boundaries, while the lower layers handle span creation, context propagation, and export. In the current implementation, thin decorators and context managers form the public API, while the span factory is responsible for creating spans, attaching the attributes in Table~\ref{tab:obs_attributes}, and linking related events across communication boundaries. This keeps the tool close to the host workflow instead of forcing users into a new orchestration model.

A further benefit of this design is that communication edges remain explicit. On send, the tool can propagate tracing context in the message carrier; on receive, it can restore that context and record the relation between the two events. This is particularly useful in agent-based workflows, where the final outcome often depends on the quality and timing of inter-agent handoffs rather than on isolated \llm calls alone.

\subsection{Fault Injection Layer}

On top of the observability layer, \tool provides a configurable fault injection layer. The guiding idea is simple: the same boundaries that are important for tracing are also the most natural points for controlled perturbation. Instead of modifying the core workflow logic, the tool evaluates configured rules at selected hook points such as \llm calls, tool invocations, A2A send, and A2A receive. If a rule matches the current execution context, the tool applies the requested action at that boundary.

This design gives the tool flexibility along two dimensions. First, faults can be targeted precisely. A specification can select a fault based on boundary type and contextual information such as phase name, acting agent, source and target agents, message channel, or similar execution metadata. Second, the same instrumentation points can support different fault behaviors, depending on the boundary. Table~\ref{tab:supported_faults} summarizes the currently supported faults in the implementation. The injection layer is also extensible: new hook types and new actions can be added without changing the host workflow code. Because the mechanism is configuration-driven, the same workflow can be re-executed under different fault scenarios without re-instrumentation.

\begin{table}[t]
\centering
\caption{Currently supported faults by injection boundary.}
\label{tab:supported_faults}
\small
\setlength{\tabcolsep}{4pt}
\renewcommand{\arraystretch}{1.08}
\begin{tabular}{p{1.35cm} p{3cm} p{3.75cm}}
\toprule
\textbf{Boundary} & \textbf{Fault} & \textbf{Meaning} \\
\midrule

A2A send &
\begin{tabular}[t]{@{}l@{}}
{\scriptsize\texttt{delay}}\\
{\scriptsize\texttt{drop}}\\
{\scriptsize\texttt{truncate}}
\end{tabular}
&
\begin{tabular}[t]{@{}l@{}}
{\scriptsize Pause before sending the message.}\\
{\scriptsize Suppress the outgoing message.}\\
{\scriptsize Shorten the message payload}\\
{\scriptsize before send.}
\end{tabular}
\\
\midrule

A2A receive &
\begin{tabular}[t]{@{}l@{}}
{\scriptsize\texttt{delay}}\\
{\scriptsize\texttt{drop}}
\end{tabular}
&
\begin{tabular}[t]{@{}l@{}}
{\scriptsize Pause before processing the}\\
{\scriptsize received message.}\\
{\scriptsize Discard the message at the}\\
{\scriptsize receive side.}
\end{tabular}
\\
\midrule

Tool call &
\begin{tabular}[t]{@{}l@{}}
{\scriptsize\texttt{delay}}\\
{\scriptsize\texttt{not\_installed}}\\
{\scriptsize\texttt{timeout}}\\
{\scriptsize\texttt{malformed\_response}}
\end{tabular}
&
\begin{tabular}[t]{@{}l@{}}
{\scriptsize Pause before tool completion.}\\
{\scriptsize Raise a tool-not-found error.}\\
{\scriptsize Raise a timeout error.}\\
{\scriptsize Return an intentionally malformed}\\
{\scriptsize tool output.}
\end{tabular}
\\
\midrule

LLM call &
\begin{tabular}[t]{@{}l@{}}
{\scriptsize\texttt{delay}}\\
{\scriptsize\texttt{rate\_limit}}\\
{\scriptsize\texttt{timeout}}\\
{\scriptsize\texttt{network\_error}}\\
{\scriptsize\texttt{malformed\_response}}
\end{tabular}
&
\begin{tabular}[t]{@{}l@{}}
{\scriptsize Pause before the model response.}\\
{\scriptsize Raise an injected rate-limit error.}\\
{\scriptsize Raise an injected timeout error.}\\
{\scriptsize Raise an injected network error.}\\
{\scriptsize Return an intentionally malformed}\\
{\scriptsize model output.}
\end{tabular}
\\

\bottomrule
\end{tabular}
\end{table}

The fault injection layer is closely tied to the trace model. When a fault injection is applied, the affected execution is still represented by the same operational span type. In other words, an A2A delay is recorded on an A2A communication span, and an \llm delay is recorded on an \llm-call span. The difference is that the span now carries fault-specific attributes and events indicating that a perturbation was injected, what type it was, which rule triggered it, and what decision was taken (Table~\ref{tab:obs_attributes}). This keeps baseline and faulty runs structurally comparable, which is essential for side-by-side analysis.

Suppose we configure the injection of a delay on the Planner$\rightarrow$Coder handoff in Figure~\ref{fig:demo_workflow}. The workflow itself does not change. What changes is that the communication boundary now carries fault metadata, and the configured delay is applied before the message continues through the workflow. In such a run, the corresponding \texttt{a2a\_send} span can still appear in exactly the same structural position as in the baseline trace, but now with attributes such as \texttt{llmmas.fault.injected=true}, \texttt{llmmas.fault.type=a2a.delay}, and \texttt{llmmas.fault.spec\_id=A2\_A2A\_DELAY}, together with scenario-specific data such as the applied delay. Similarly, if the injected fault targets the Planner's \llm call, the planning-phase \texttt{llm\_call} span remains in place, but it is annotated as faulted and delayed. This is an important design decision of \tool: fault injection is not treated as a separate logging mode or a disconnected experiment pipeline, but as an overlay on the same execution structure used for normal runs.

Taken together, the observability and fault injection layers provide a single experimental setup for debugging and evaluation. Observability gives a structured account of what happened during a run. Fault injection introduces controlled perturbations at explicit boundaries. It helps users answer concrete engineering questions about their workflows: which boundaries are most sensitive to perturbation, whether a local issue remains localized or cascades across later phases, and whether the workflow degrades through delay, divergence, retries, or outright failure. Combined, they support a more systematic analysis of \llmmas behavior than final outputs or flat logs alone.

%% file: content/03_validation.tex
\section{Validation}
\label{sec:validation}

In this section, we demonstrate how \tool can be used to run controlled stress test scenarios and quantify their runtime effect. We use the same setup across two target systems: the minimal Planner$\rightarrow$Coder workflow used throughout this paper, and an existing LLM-MAS. In both cases, we use the 30-task ProgramDev benchmark ~\cite{cemri2025mast} and execute each task five times per condition. We consider three scenarios: a fault-free baseline, a run with a single \texttt{llm.delay = 1000 ms} injected at the first \llm call in the planning phase, and a run with a single \texttt{a2a.delay = 1000 ms} injected at the Planner$\rightarrow$Coder send in the planning phase. To summarize the runtime effect of an injected fault, we use \emph{amplification}, defined as the extra end-to-end runtime caused by the fault divided by the injected delay. An amplification close to 1 means that a 1000\,ms injected delay produces roughly a 1000\,ms slowdown overall, while values above 1 indicate that the injected delay propagates and causes a larger slowdown than the perturbation itself.

We first applied this setup to the demo workflow. Table~\ref{tab:validation_results} shows that both injected scenarios lead to measurable end-to-end slowdown. In the demo, injecting the planning-phase \llm delay yields a mean amplification of 1.053 and a median of 1.036, which is close to linear overhead. In contrast, delaying the Planner$\rightarrow$Coder handoff yields a higher mean amplification of 1.295 and a median of 1.390, suggesting that perturbing the inter-agent communication boundary causes stronger downstream slowdown than delaying the single \llm call alone.

For the second setting, we use ChatDev ~\cite{qian2024chatdev}, a chat-powered software development framework in which multiple specialized software agents collaborate through language-based communication across three main software lifecycle phases: design, coding, and testing. It is therefore a useful second validation target because it represents a richer and more realistic system than the minimal demo, while still remaining firmly in the software engineering domain. Table~\ref{tab:validation_results} shows that delay faults in ChatDev lead to substantially higher amplification factors, reflecting how a localized perturbation can cascade across many downstream interactions and \llm calls. In particular, injecting a single \llm delay yields a mean amplification of 48.1 and a median of 13.9, while delaying an inter-agent message boundary yields an even higher mean amplification of 59.2 (median 6.6). Overall, these results highlight that in a multi-phase, message-heavy workflow, the runtime impact of small injected delays can be magnified dramatically, and the proposed tracing+injection setup makes this amplification visible and quantifiable in a trace-aligned manner.

\begin{table}[t]
\centering
\caption{Amplification results for the validation scenarios.}
\label{tab:validation_results}
\small
\begin{tabular}{llccc}
\toprule
\textbf{System} & \textbf{Scenario} & \textbf{Mean} \\
\midrule
Demo & Baseline & -- \\
Demo & LLM delay & 1.053 \\
Demo & A2A delay & 1.295  \\
\midrule
ChatDev & Baseline & -- \\
ChatDev & LLM delay & 48.1 \\
ChatDev & A2A delay & 59.2 \\
\bottomrule
\end{tabular}
\end{table}

%% file: content/04_related_work.tex
\section{Related Work}
\label{sec:related}

A line of work focuses on \emph{understanding} and \emph{classifying} failures in multi-agent systems. MAST provides the first empirically grounded taxonomy of multi-agent \llm system failures, organizing failure modes around specification issues, inter-agent misalignment, and task verification problems~\cite{cemri2025mast}. AgentFail extends this direction to platform-orchestrated agentic systems by proposing a root-cause taxonomy across agent-, workflow-, and platform-level failures, and by releasing a benchmark for diagnosing these causes~\cite{ma2025diagnosing}. Related work, such as TRAIL studies, structured agent traces for issue localization and debugging, while Who\&When focuses on identifying the decisive agent and step responsible for a failed outcome~\cite{deshpande2025trail,zhang2025whowhen}. These papers are highly relevant to our motivation: they show that failures in \llmmas are diverse, often distributed across long traces, and difficult to localize by manual inspection alone. However, their main contribution is post-hoc analysis, attribution, or taxonomy building. They do not provide a general-purpose tool layer for injecting controlled perturbations into live executions and comparing baseline and faulty runs through a shared trace structure.

Another line of work is more directly related to fault injection. AEGIS~\cite{kong2025aegis} generates faulty multi-agent trajectories by taking successful executions and applying controlled, context-aware error injections, producing labeled data for attribution and analysis. AgenTracer~\cite{zhang2025agentracer} also uses synthetic failure generation, combining counterfactual replay with programmatic perturbation of selected steps so that the decisive faulty step is known by construction. These are the closest works to \tool, because they treat controlled perturbation as a way to study failure. Still, their primary goal is dataset construction and failure attribution rather than providing a reusable runtime tool for engineering evaluation. In AEGIS and AgenTracer, injection is mainly a means to generate faulty trajectories for learning or labeling. In contrast, \tool packages fault injection as a configurable execution-time capability around an existing workflow. The goal is not only to create faulty variants, but also to support repeatable stress-testing, baseline-versus-faulty comparison, and trace-aligned diagnosis at concrete boundaries such as \aTwoA communication, tool calls, and \llm calls.

A third line of work addresses observability and operations for agentic systems. LumiMAS~\cite{solomon2025lumimas} argues for real-time monitoring of multi-agent systems and highlights security- and reliability-relevant phenomena such as prompt injection, memory poisoning, and cascading hallucinations. AgentOps~\cite{dong2024agentops} similarly frames observability as a core requirement for autonomous and nondeterministic agents, and discusses the artifacts and telemetry that should be captured across the agent lifecycle. More broadly, recent surveys of \llmmas emphasize persistent challenges around robustness, memory, coordination, tool use, and evaluation~\cite{li2024survey}. These works are important for our observability layer, as they motivate why traces must capture more than final outputs. At the same time, they stop short of coupling observability with controlled perturbation. In our setting, observability is not an independent monitoring dashboard; it is the mechanism that makes injected runs interpretable and comparable.

Overall, existing work provides three important ingredients: taxonomies of what can go wrong, attribution methods for identifying where it went wrong, and observability proposals for capturing what happened. What is still missing is a lightweight tool that brings these ideas together for controlled experimentation on existing \llmmas, especially in software engineering workflows. \tool is designed to fill that gap. 
It provides the practical infrastructure to run repeatable baseline and fault-injected executions, capture them in a shared trace model, and inspect how faults propagate across phases, agents, and interaction boundaries.

%% file: content/05_limitations_future_work.tex
\section{Limitations and Future Work}
\label{sec:limitations}

\tool provides a modular and extensible foundation for stress testing \llmmas. The current implementation supports fault injection at three key runtime boundaries (agent-to-agent communication, \llm calls, and tool calls), which already enables controlled perturbation of important interaction points while preserving aligned traces for baseline and injected runs. Ongoing work extends this capability toward additional fault types and broader injection surfaces, including higher-level workflow, memory-related, and coordination faults.

The current version of \tool focuses on instrumentation, trace-aligned artifact generation, and configurable runtime perturbation. It does not yet provide higher-level analysis features such as paired-run differencing, trace summarization, root-cause ranking, or automated debugging reports; at present, users inspect the produced traces and run artifacts directly. A next step is therefore to build analysis support on top of the current observability layer.

%% file: content/06_tool_availability.tex
\section{Tool Availability}
\label{sec:availability}

\tool is publicly available, along with a sample demo, ChatDev instrumentation, and detailed instructions, at \href{https://github.com/vagabondboffin/llmmas-otel}{GitHub}.

%% file: content/07_conclusion.tex
\section{Conclusion}
\label{sec:conclusion}

We presented \tool, a lightweight and framework-agnostic tool that combines structured observability with fault injection for \llmmas in software engineering. By keeping baseline and injected executions aligned under a shared trace model, it enables controlled stress testing and systematic inspection of how faults affect multi-agent workflows. The tool can serve as a practical foundation for future work on the reliability and evaluation of \llmmas. 

%% file: content/08_acknowledgment.tex
\section*{Acknowledgment}
This work was conducted as part of the AI for Software Engineering (AI4SE) collaboration between JetBrains and Delft University of Technology. The authors gratefully acknowledge the financial support provided by JetBrains, which made this research possible.